\documentclass[prl,twocolumn,floatfix,superscriptaddress,showpacs]{revtex4}
\usepackage{graphicx,dcolumn,bm}
\newcommand{\tHe}{$^3\rm He$}

\begin{document}

\newcommand{\FOCUS}{FOCUS Center and Physics Department, University of Michigan,
Ann Arbor, MI 48104, USA}
\newcommand{\ILL}{Institut Laue Langevin, BP 156, 38042 Grenoble Cedex 9, France}
\newcommand{\HD}{Physikalisches Institut, Universit{\"a}t Heidelberg, Philosophenweg 12, 69120 Heidelberg, Germany}
\newcommand{\ISIS}{ISIS, Rutheford Appelton Labs, Chilton, Didcot OX11 0QX, UK}
\newcommand{\NIST}{National Institute of Standard and Technology,Gaithersburg, MD 20899-8461, USA}
\newcommand{\UNAM}{Universidad Nacional Aut\'{o}noma de M\'{e}xico, M\'{e}xico, D.F.
04510, M\'{e}xico}
\newcommand{\LANL}{Los Alamos National Lab, Los Alamos, NM  87545 USA}
\newcommand{\ORNL}{Oak Ridge National Lab, Oak Ridge, TN 37831 USA}
\newcommand{\ASU}{Arizona State University, Tempe, AZ 85287 USA}
\newcommand{\UW}{University of Wisconsin, Madison, WI 53706 USA}

\title{Neutron Beam Effects on 
 Spin-Exchange-Polarized $^3$He}

\author{M. Sharma*}\affiliation{\FOCUS} 
\author{E. Babcock} \affiliation{\ILL} 
\author{K.H. Andersen} \affiliation{\ILL} 
\author{L. Barr\'{o}n-Palos}\affiliation{\ASU} \affiliation{\UNAM}
\author{M. Becker} \affiliation{\ILL}\affiliation{\HD} 
\author{S. Boag} \affiliation{\ISIS} 
\author{W.C. Chen}\affiliation{\NIST}
 \author{T.E. Chupp}  \affiliation{\FOCUS} 
 \author{A. Danagoulian}\affiliation{\LANL}
 \author{T.R. Gentile}\affiliation{\NIST} 
  \author{A. Klein}\affiliation{\LANL} 
 \author{S. Penttila}\affiliation{\ORNL}
 \author{A. Petoukhov}\affiliation{\ILL}
\author{T. Soldner} \affiliation{\ILL} 
\author{E.R. Tardiff}\affiliation{\FOCUS}
 \author{T.G. Walker}\affiliation{\UW}
 \author{W.S. Wilburn}\affiliation{\LANL}

 \date{\today.  To be submitted to Phys. Rev. Lett.}

\begin{abstract}
We have observed depolarization effects when  high intensity cold neutron
beams are incident on alkali-metal-spin-exchange polarized $^3$He cells
used as neutron spin filters. This was first observed as a reduction of 
the maximum attainable $^3$He polarization and was attributed to a decrease of alkali-metal polarization, which led us to  directly measure alkali-metal polarization and spin relaxation  over a range
of neutron fluxes at LANSCE and  ILL. The data reveal a new alkali-metal spin-relaxation mechanism that  approximately scales as $\sqrt\phi_n$,
where $\phi_n$ is the neutron capture-flux density incident on the cell. This is
consistent with an effect proportional to the recombination-limited ion concentration, but is much larger than expected from earlier work. 
\end{abstract}

\pacs{32.80.Xx, 24.70.+s, 03.75.Be, 61.80.Hg}
\maketitle
Polarized gaseous $^3$He has wide application including targets and
beams for nuclear physics measurements,
for electron
scattering studies of the structure of the
neutron\cite{rf:ChuppMilnerHolt}, for biomedical imaging of the airspace in
the lungs\cite{rf:Kauzor}, and as a neutron
polarizer\cite{rf:Coulter89,rf:Chupp07,rf:someILL}. Each of these
applications has a different set of requirements and acceptable
tradeoffs of polarization, density, size and polarization stability.
There are two techniques used to produce polarized $^3$He gas: metastability exchange
optical pumping (MEOP) and spin-exchange optical pumping (SEOP).
MEOP polarizes pure $^3$He at low pressure, typically 1 mbar, at rates of about 1 bar-liter/hour with $^3$He polarizations of 70\% or more\cite{rf:someILL}. MEOP polarizer stations  compress the $^3$He into cells that are transported to the point of use where the $^3$He polarization
decays very slowly, with a time constant that can be a week or longer\cite{rf:someILL2}. 
 For SEOP, the $^3$He is polarized by the hyperfine interaction during collisions of the $^3$He
nuclei with polarized valence electrons of optically pumped alkali-metals. Production rates with
SEOP are about an order of magnitude lower than the highest MEOP
rates but similarly high $^3$He polarizations have been acheived\cite{HybridCells}.
For applications that require several days or weeks of stable high
polarization operation, such as targets for electron scattering, neutron
scattering instruments with limited access, and long-running fundamental neutron
physics experiments,  it is practical to have a SEOP system pumping continuously, with stable polarization, for weeks or months \cite{rf:Chupp07,rf:JRJohnson}.

Gaseous polarized $^3$He is used for polarized neutron measurements because
of the nearly complete spin-dependence of  the absorption cross
section for the process $^3$He(n,p)$^3$H+764 keV. This
 proceeds through an unbound $0^+$ resonance in $^4$He,  so that only neutrons with spin opposite to the $^3$He
spin are absorbed\cite{rf:Passel66}. 
The effective absorption cross section for neutrons of wavelength $\lambda$ can be written
\begin{equation}
\sigma_a= \sigma_0{\lambda\over \lambda_0} {1\mp P_{3}\over 2}, 
\end{equation}
for neutron spin parallel ($-$) or antiparallel ($+$) to the $^3$He polarization.
Here $\sigma_0= 5333\pm 7$ b is the  absorption cross section for thermal neutrons ($\lambda_0$ = 1.8 \AA)\cite{rf:Mughabghab}, and $P_{3}$ is the magnitude of the
$^3$He polarization.
In a spin-filter polarizer, the $^3$He polarization is not complete, and neutrons of both spin states are absorbed, though
with different absorption lengths. Thus the wavelength dependent transmission and
polarization of the neutrons are given by
$$
T_n(\lambda)=e^{-\sigma_0 t_3{\lambda\over \lambda_0}}\cosh(P_{3}\sigma_0 t_3{\lambda\over \lambda_0}),
$$
\begin{equation}
\label{eq:PnTn}
P_n(\lambda)=\tanh(P_{3}\sigma_0 t_3{\lambda\over \lambda_0}),
\end{equation}
where
 $ t_{3}$ is the $^3$He
areal density. Transmission of polarized neutrons through polarized  $^3$He can also be used to analyze the neutron polarization with an analyzing power $A_n=\tanh(P_{3}\sigma_0 t_3{\lambda\over \lambda_0})$.

In a typical SEOP neutron spin filter, the $^3$He cells are constructed from boron-free glass, {\it e.g.} GE180\cite{rf:NISTHe3}.  Alkali-metal (rubidium\cite{rf:Chupp87} or a mixture of rubidium and potassium\cite{Hybrid}) is distilled into the cell with about one bar of \tHe\ at room temperature and a small
amount of N$_2$ added to suppress radiation trapping, the multiple scattering of optical pumping photons that depolarize the alkali-metal atoms\cite{rf:Chupp85}. The cell is heated to maintain an optimum alkali-metal vapor pressure, held in a magnetic field of 10-30 Gauss, and illuminated by a high powered laser tuned to the rubidium D1 resonance at 794.7 nm.
The $^3$He polarization is governed by an exponential time dependence with rate constant and equilibrium polarization  given respectively by
\begin{equation}
\Gamma=(1+X_{cell})\gamma_{SE}+\Gamma_R
\quad
\quad
P_3^{eq}=P_{A}{\gamma_{SE}\over\Gamma}
\end{equation}
where $\gamma_{SE}=\langle \sigma_{SE}v\rangle[\textrm {Rb}]$, the velocity-averaged spin-exchange rate constant, is typically $1/(10-15\ \textrm h)$.
The  $^3$He relaxation rate, $\Gamma_{R}$,  is a sum of rates due to cell
wall interactions, impurities, \tHe\ dipole-dipole relaxation, magnetic field gradients and ionization effects. The rate $\Gamma_R$ is generally 10-50 times smaller than $\gamma_{SE}$.
The volume averaged alkali metal electron polarization is $P_A$, and
the factor $X_{cell}$ accounts for an observed reduction
in \tHe\ polarization that varies from cell-to-cell  \cite{rf:Babcock2006}.

\begin{figure}
\includegraphics[width=3.3 truein]{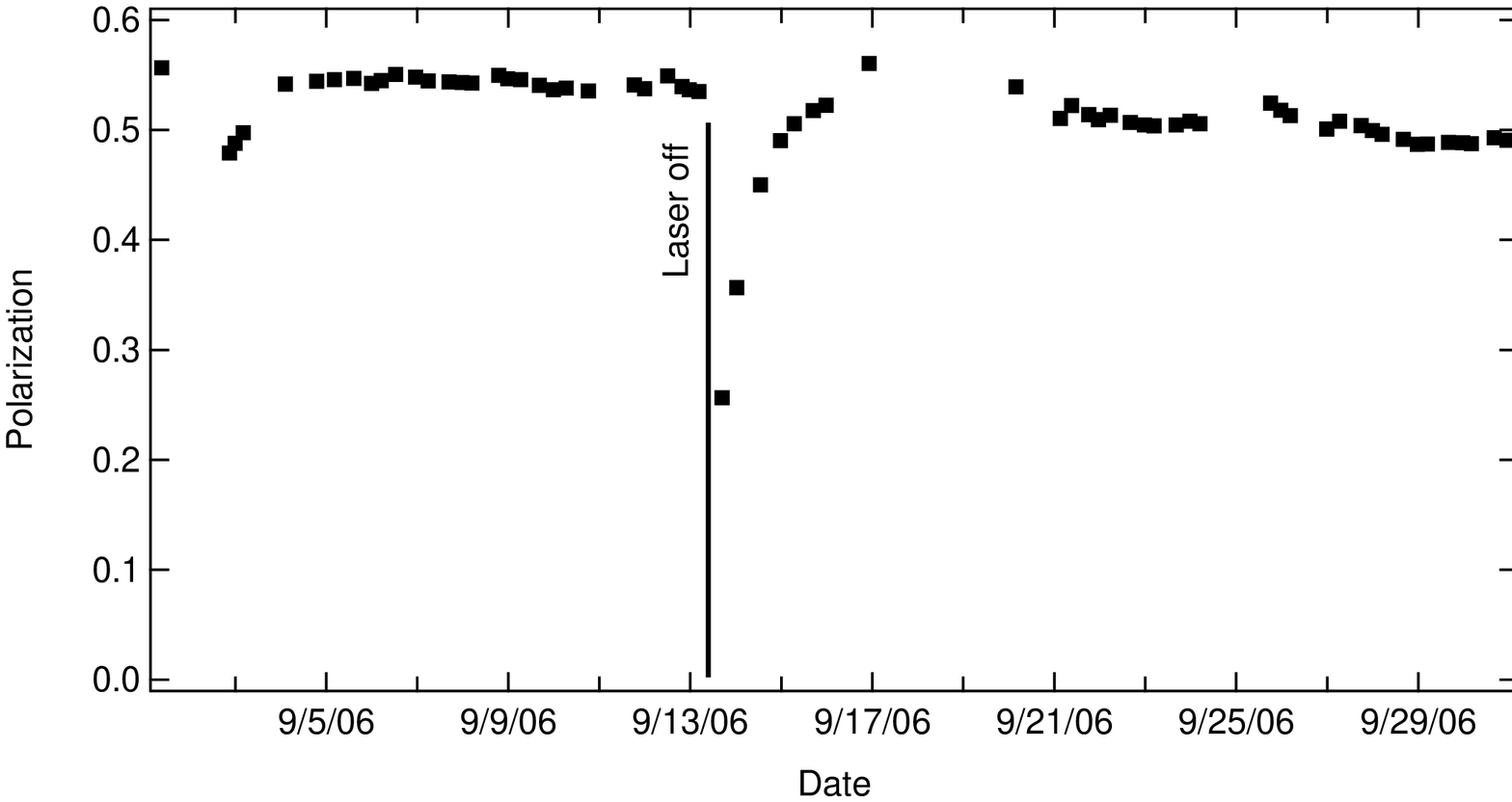}
\includegraphics[width=3.3 truein]{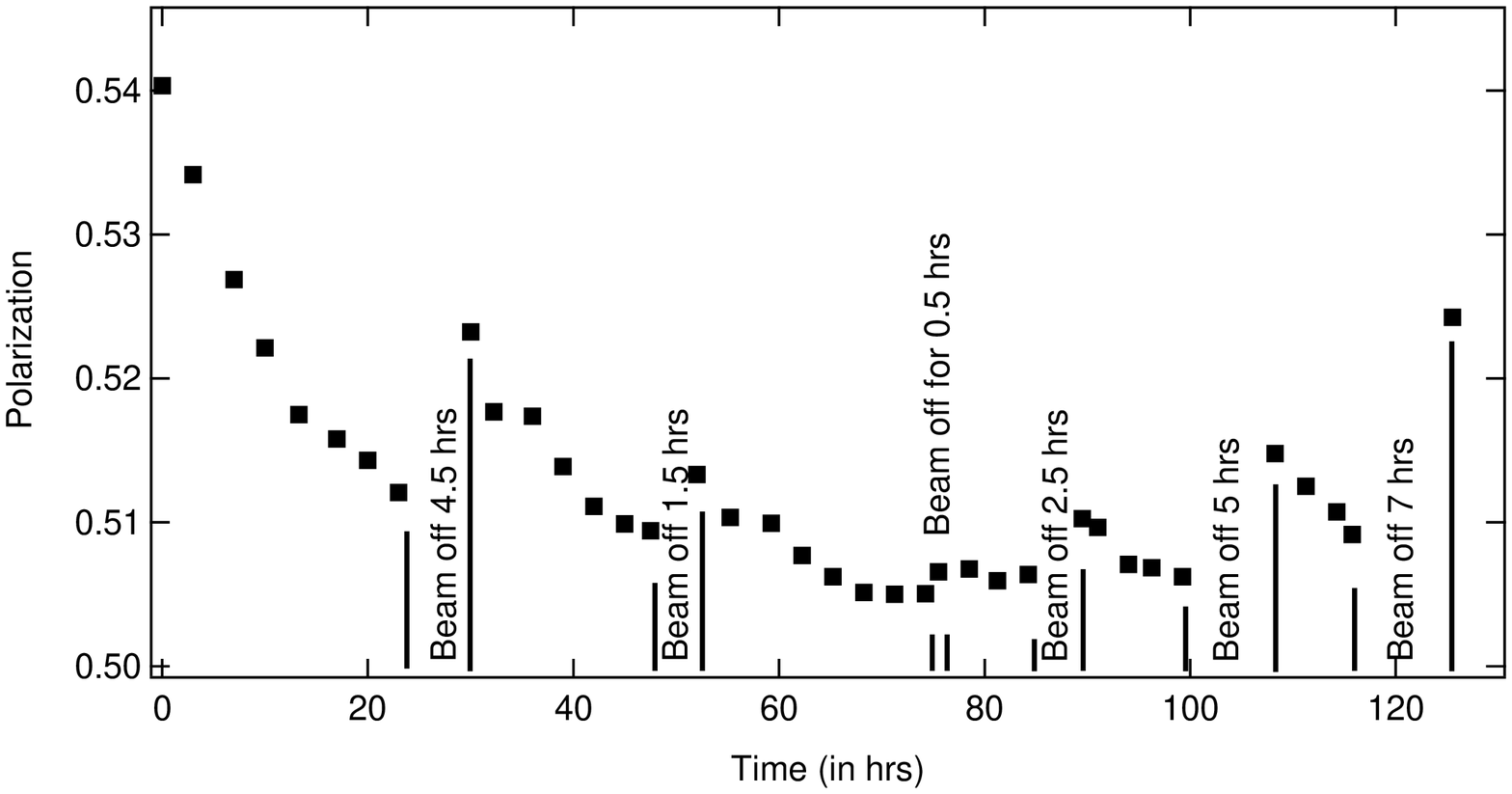}
\caption{LANSCE neutron spin filter $^3$He polarization. The top panel shows the long term behavior; the bottom panel shows the effect of the neutron beam
on short time scales.}\label{poldrop}
\end{figure}

Neutron beam effects on the $^3$He polarization were first observed
during development of the 
NPDGamma
experiment at
 the Los Alamos Neutron Science Center (LANSCE)\cite{rf:Gericke2007,rf:Chupp07} and were further studied in 
dedicated runs at LANSCE 
and
at the
Institute Laue-Langevin (ILL) in Grenoble.
 The $^3$He polarization for a cell used at LANSCE over two
months 
is shown in Figure \ref{poldrop}. The cells and set-up are described in reference \cite{rf:Chupp07}. The top panel shows that, though the $^3$He
polarization appears relatively constant over the long term (except for the period with
the laser off), there is a slow downward drift. The long
time constant decay of $^3$He polarization appears to be due to a milky white coating that builds up on the cell walls and reduces transmission of  laser light into the cell\cite{rf:Chupp07}. This build-up is probably due to
rubidium compounds, 
possibly due to reaction
with the hydrogen ($^1$H and $^3$H) produced by neutron absorption on $^3$He. 
A similar effect was observed for a pure rubidium  cell at 170$^\circ$C placed in the full flux PF1B beam at ILL for one hour.
Exposure for one hour at PF1B is equivalent to about three days at LANSCE FP12. A Monte Carlo calculation based on the measured brightness of the LANSCE neutron source \cite{rf:Pil} predicts a maximum capture-flux density of
$1-3\times 10^{8}$ cm$^{-2}$s$^{-1}$. (Capture flux, the integral of the $1/v$ weighted neutron intensity spectrum, is proportional to the total neutron capture or decay rate per unit length.)
The PF1B capture-flux density at the cell position was measured with gold foils and found to be $1.4\times 10^{10}$ cm$^{-2}$s$^{-1}$. The PF1B beam is described in reference \cite{rf:Abele2006}

Data for $P_3$ on shorter time scales are shown in the bottom panel of Figure 1.  When the  beam is on, the $^3$He polarization decays, and with the beam off, the polarization recovers, at least partially. 
The short time-scale data of Figure \ref{poldrop} show that  the neutron beam causes the $^3$He polarization to decay to a lower value of $P_3^{eq}$ at a rate of approximately (1/12 h), which is consistent with the measured $\Gamma$.  
The polarization was not measured with the neutron beam off, because we used the neutrons to measure $P_{3}$ \cite{rf:Chupp07}; however the increase of $^3$He polarization is consistent with a similar rate constant.
Since $\Gamma$ does not change appreciably, the most likely cause
is a  drop of $P_{A}$, possibly due to ionization effects induced by the neutron beam.
Ionization effects
on both $\Gamma$ and $P_A$ were observed in work with a 180 particle-nA beam of 18 MeV alpha particles\cite{rf:Coulter91}. Those observations led to the development of the double cell now ubiquitous in SEOP  based polarized $^3$He targets for electron scattering \cite{rf:Chupp92}; however we expected these effects to be negligible for $^3$He cells in  neutron beams, where the ionization energy loss is  100 to 10,000 times less.  We therefore set out to measure the effects of the neutron beam on the alkali-metal polarization in high-flux neutron beams under a range of conditions.

\begin{figure}
\includegraphics[width=3 truein]{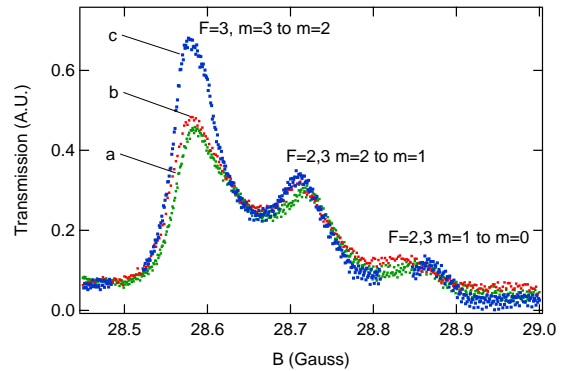}
\caption{ESR spectra of $^{85}$Rb for full flux (a: green) and 19\% of full flux (b: red) and no beam (c: blue) at LANSCE. The polarizations are $P_A$ =  0.6, 0.64 and 0.7, respectively. 
The relative uncertainties, estimated to be 1-2\%, are limited by the signal-to-noise ratio of the ESR measurements.}
\label{LANSCEesr}
\end{figure}

The volume averaged alkali-metal polarization was directly measured using
electron-spin resonance (ESR)\cite{ESRPaper} at both LANSCE and ILL. In the LANSCE set-up 
the transmission of optical pumping light from a single 30 W broadband laser-diode array was
monitored as RF at 14 MHz was applied and the magnetic field swept from 28.4 to 29.0 Gauss. The magnetic field
was produced by the combination of a  very uniform 10 Gauss field
from a large set of coils and a pair of hand-wound 
rectangular coils that produced a 
less homogeneous  field of about 20 Gauss. 
The
consequence of the inhomogeneity 
is that the ESR lines are
broadened so that the hyperfine lines are not all separately resolved.
Data for two different neutron beam intensities and no beam are shown in Figure \ref{LANSCEesr}. When the hyperfine levels are not resolved, the rubidium polarization is given by $P_{A}=(7R-3)/(7R+3)$ for $^{85}$Rb ($I=5/2$),
where $R$ is the ratio of adjacent ESR peak areas extrapolated to zero RF power.

\begin{figure}
\includegraphics[width=3. truein]{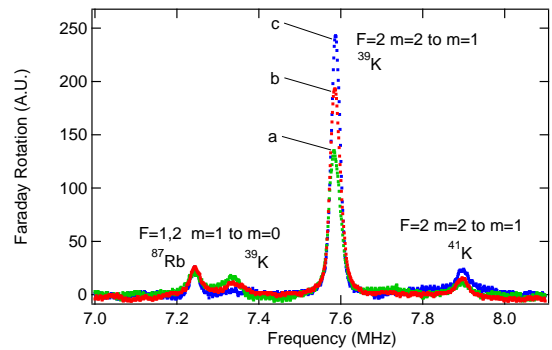}
\caption{ESR Spectra of $^{39,41}$K and $^{87}$Rb ESR for full flux (a: green) and 8.5\% of full flux (b: red) and no flux (c: blue) at ILL. The polarizations are $P_A$ =  0.83, 0.90, 0.98, respectively. }\label{OrvietoESR}
\end{figure}

For the ILL set up, the ESR measurements were made at 10 Gauss with
a hybrid Rb-K cell constructed at NIST \cite{HybridCells}. 
The second-order Zeeman splitting of the potassium is much larger than that of rubidium and
allows  ESR  of $^{39}$K to be resolved at
10 Gauss as shown in Figure \ref{OrvietoESR}. The signals from
$^{87}\rm{Rb}$, and
$^{41}\rm{K}$\cite{Hybrid} (both $I=3/2$) are also observed. Rapid spin
exchange between the rubidium and potassium\cite{WilmerSpinExch} allows the ESR of
potassium to measure the average
electron polarization of all the alkali-metal  species.  
The cell was illuminated with light from two 100 Watt narrowed diode-laser-array bars\cite{WiscNarrowBars}.
A  linearly polarized probe laser, tuned near
the Rb D2 resonance and directed along the magnetic field, was used to measure the Faraday rotation signal, which is proportional to the alkali-metal polarization along the probe beam path through the vapor\cite{faraday}. The RF field was swept over a range of approximately 1.1 MHz around  7.6 MHz and the data
extrapolated to zero RF power. For $I=3/2$ $P_{A}=(2R-1)/(2R+1)$. 

\begin{figure}
\includegraphics[width=3 truein]{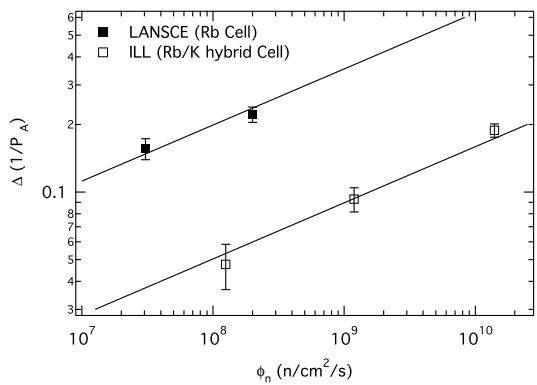}
\caption{Neutron induced change in polarization for the LANSCE Rb cell and  ILL hybrid cell. The solid lines are provided to guide the eye.}
\label{fg:DeltaP}
\end{figure}

The alkali-metal polarization at any position in the cell, is given by\cite{rf:Chupp87}
\begin{equation}
{1\over P_{A}(\vec r)}= { 1+ {\Gamma_{SD}\over \gamma_{opt}(\vec r)}},
\label{eq:PA}
\end{equation}
where  $\gamma_{opt}(r)$  is the convolution of the laser spectral profile and the optical absorption cross section at the position $\vec r$. The spin destruction rate, $\Gamma_{SD}$, is the rate of electron spin-flips per alkali-metal atom and most likely changes more significantly than 
$\gamma_{opt}(\vec r)$.
In Figure \ref{fg:DeltaP}   we plot the change  $\Delta(1/P_A)$ relative to no beam
as a function of neutron capture-flux density, $\phi_n$, for both the LANSCE  and ILL data.  With the 
higher power, narrowed lasers pumping the hybrid cell at ILL, the neutron-beam effects are significantly reduced  compared to the LANSCE data at a given neutron flux density.

\begin{figure}
\hskip -0.05 truein
\includegraphics[width=3.1 truein]{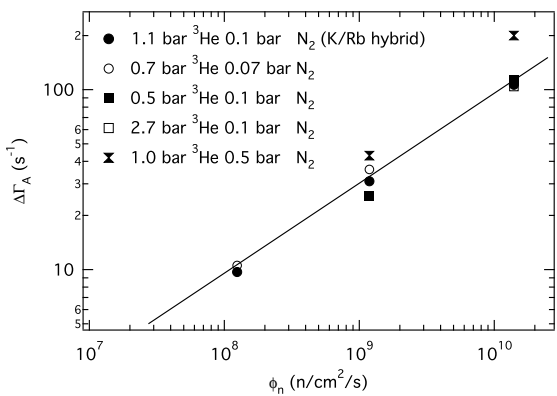}
\caption{ILL polarization relaxation results for r all cells. Error bars of $\approx$10\% are similar to the size of the symbols. 
The solid line, provided to guide the eye, is proportional to  $\sqrt{\phi_n}$.}\label{results}
\end{figure}

Relaxation of the alkali-metal polarization  was studied at ILL using the  relaxation
 in the dark technique \cite{RelaxationintheDarkRef}.  A small alkali-metal polarization was produced by a low power  optical pumping beam  (less than 0.1W/cm$^2$), which was chopped at 1 Hz. The polarization,   $P_A$,  was measured by Faraday rotation with the same setup used for the data shown in Figure \ref{OrvietoESR}. With the optical-pumping beam chopped off, the polarization decayed  at a rate $\Gamma_A=\Gamma_{SD}/S$, where the slowing factor  $S\ge 1$  accounts for the angular momentum stored in the nuclear spins, which couple to the electron spin through the hyperfine interaction\cite{rf:Wagshul94}. Due to electron spin-exchange, the factor $S$ is an average over isotopes and alkali-metal species. The slowing factor depends on the alkali-metal polarization; for low polarization,  $S=10.8$ for natural rubidium\cite{rf:Wagshul94}, and $S=6$ for potassium\cite{Hybrid}.
Results for  $\Delta\Gamma_A$,  the neutron-flux contribution to the relaxation rate, for all five cells measured at ILL, each  with different gas and alkali-metal compositions, are shown in Figure  \ref{results}. With no beam, $\Gamma_A$ varies from 20 to 30 s$^{-1}$ depending on gas and alkali-metal compositions and pressures. 
The solid line in Figure 5 has the form $\Delta\Gamma_{A}\propto \sqrt{\phi_n}$. As shown below, this would be consistent with relaxation due to a recombination-limited equilibrium ion concentration.

The processes due to the ionization, created mainly by the $^3$He(n,p)$^3$H reaction, are complex and involve ions, metastable $^3$He atoms, molecular ions and radicals of helium and nitrogen. One or more of these species may be the cause of the observed effects.
 We consider 
 the simplified case of production of an ion species of density $n_i$ at a rate proportional to the neutron capture-flux density $\phi_n$:
${dn_i\over dt }= \gamma\phi_n - \beta n_i$, where $\gamma$=(764 keV/$\delta E_i)/L$ for a cell of length $L$, $\delta E_i$ is the energy cost per ion-pair ({\it e.g.} $\delta E_i$=32.5 eV for helium),
and $\beta$ is the rate constant for recombination due all mechanisms including electron recombination, neutralization at the cell wall, and charge exchange with species  that do not contribute to $\Delta P_A$ ($\beta = \beta_e + \beta_W + \beta_{ex} +\dots $).
In the limit that electron recombination is dominant, the equilibrium electron density will be proportional to $n_i$, {\it i.e.} $\beta_e=\alpha n_i$, and 
the equilibrium solution is, $n_i = \sqrt{\gamma\phi_n/\alpha}$, consistent with the solid line in Figure 5. 

In summary, measurements at LANSCE and ILL with in-situ SEOP $^3$He neutron spin filters have shown that the incident neutron beam  induces  an decrease of alkali-metal polarization and a corresponding  increase of the alkali-metal relaxation rate. Measurements over several decades of neutron flux show that the increased spin relaxation rate approximately scales with  $\sqrt{\phi_n}$,  which would be consistent with
the  recombination-limited equilibrium concentration of one or more ion species.
The magnitude of this effect is much larger than expected given earlier study of  ionization effects produced by an alpha-particle beam\cite{rf:Coulter91}.  At ILL's PF1B, the world's highest flux cold neutron beam for fundamental physics, $P_A$ was reduced by about 20\% in a potassium-rubidium hybrid cell pumped by high-power narrowed diode laser arrays. 
 Further neutron-beam related effects were observed in the ILL measurements including
 a combination of prompt and delayed changes in the alkali-metal relaxation rates,  cell pressure dependent effects and  performance of a double cell. These will be presented in a separate paper. 
 
\begin{acknowledgments}
This work was supported by the U. S. National Science Foundation, the Department of Energy, the ILL Millennium Programme, and the NMI3. We gratefully acknowledge the efforts of the entire NPDGamma collaboration in developing the apparatus used to provide the data shown in Figure 1, and we gratefully acknowledge the technical assistance of the ILL $^3$He group.
\end{acknowledgments}

\end{document}